\documentclass[mathleft
]{an}
\usepackage{graphicx}
\usepackage{times}
\newcommand\E[1]{\times10^{#1}}
\newcommand\parfrac[2]{\left(\frac{#1}{#2}\right)}
\newcommand{\simgt}{\lower.5ex\hbox{$\; \buildrel > \over \sim \;$}}
\newcommand{\simlt}{\lower.5ex\hbox{$\; \buildrel < \over \sim \;$}}

\newcommand\U[1]{{\,\rm #1}}

\newcommand\bt{\beta}
\newcommand\gm{\gamma}
\newcommand\Gm{\Gamma}
\newcommand\eps{\epsilon}
\newcommand\sg{\sigma}
\newcommand\Lmb{\Lambda}
\newcommand\rs[1]{_\mathrm{#1}}
\newcommand\LX{L\rs{X}}
\newcommand\Edot{\dot E}
\newcommand\tsyn{t\rs{syn}}
\newcommand\epssyn{\eps\rs{syn}}
\newcommand\ESN{E\rs{SN}}
\newcommand\Mej{M\rs{ej}}
\newcommand\Edotz{\dot E\rs{o}}
\newcommand\tauz{\tau\rs{o}}
\newcommand\etae{\eta_e}
\newcommand\etaB{\eta_B}
\newcommand\etaP{\eta_P}
\newcommand\nISM{n\rs{ISM}}
\newcommand\rhoISM{\rho\rs{ISM}}
\newcommand\Vz{V\rs{o}}
\newcommand\Phiz{\Phi\rs{o}}
\newcommand\tc{t\rs{c}}
\newcommand\tcm{t\rs{c-}}
\newcommand\tcp{t\rs{c+}}
\newcommand\RPWN{R\rs{PWN}}
\newcommand\BPWN{B\rs{PWN}}
\newcommand\nusyn{\nu\rs{syn}}
\newcommand\tad{t\rs{ad}}
\newcommand\RPWNdot{\dot\RPWN}
\newcommand\teq{t\rs{eq}}
\newcommand\Lsyn{L\rs{syn}}
\newcommand\GmPWN{\Gm\rs{PWN}}
\newcommand\Msun{M_\odot}

\begin{document}

\Pagespan{1}{}
\Yearpublication{2013}%
\Yearsubmission{2013}%
\Month{11}%
\Volume{999}%
\Issue{88}%

\title{Modeling statistical properties of the X-ray emission from aged Pulsar Wind Nebulae}

\author{R. Bandiera\inst{1}\fnmsep\thanks{Corresponding author:
  \email{bandiera@arcetri.astro.it}\newline}
}
\titlerunning{Statistical properties of aged PWNe}
\authorrunning{R. Bandiera}
\institute{
INAF - Osservatorio Astrofisico di Arcetri, Largo E. Fermi 5, I-50125 Firenze, Italy}

\received{15 Sep 2013}
\accepted{xx Xxx 2013}
\publonline{later}


\abstract{%
The number of known PWNe has recently increased considerably, and the majority of them are now middle-age objects.
Recent studies have shown a clear correlation of both X-ray luminosity and size with the PWN age, but fail in providing a thorough explanation of the observed trends.
Here I propose a different approach to these effects, based on the hypothesis that the observed trends do not simply reproduce the evolution of a ``typical'' PWN, but are a combined effect of PWNe evolving under different ambient conditions, the leading parameter being the ambient medium density.
Using a simple analytic approach, I show that most middle-age PWNe are more likely observable during the reverberation phase, and I succeed in reproducing trends consistent with those observed, provided that the evolution of the X-ray emitting electrons keeps adiabatic over the whole reverberation phase.
As a direct consequence, I show that the X-ray spectra of older PWNe should be harder: also this is consistent with observations.
}
\maketitle

\section{Introduction}
Pulsar Wind Nebulae (hereafter PWNe) are bubbles of magnetic fields and relativistic particles (see e.g.\ Gaensler \& Slane 2006). They are powered by the relativistic wind of a spinning-down pulsar, and show up as powerful synchrotron (+ inverse Compton) sources.
In the past these nebular sources were dubbed as ``plerions'', from their ``filled-center'' radio morphology  (Weiler \& Panagia 1978), and relatively few objects were known in radio (Weiler 1983), as well as in the X-rays (Seward 1989).
A general idea at that time was that plerions were ``short-lived'' sources, with typical ages up to about $\sim10^4\U{yr}$, comparable with the initial pulsar spin-down time ($\tauz$), namely the time over which most of the initial pulsar rotational energy is released and converted into magnetic fields and relativistic particles.

This paradigm has changed completely in very recent years, after the identification as PWNe of many Galactic sources (Hinton \& Hofmann 2009) discovered in the TeV range, mainly with the H.E.S.S. telescope.
Follow-up campaigns in the X-rays (for the nebula) and in radio (for the pulsar) have allowed a characterization of many of these objects.
There are now about 70 Galactic PWNe known.
A very recent list can be found in Kargaltsev, Rangelov \& Pavlov~(2013): in this list PWNe are typically ``aged'' sources, with ages in the range $10^4$--$10^6\U{yr}$, much more than originally expected; moreover, the increased number of sources justifies serious attempts to derive statistical properties of the sample.

In this paper I will mainly focus on two recently measured correlations: that between the PWN X-ray luminosity ($\LX$) and the present pulsar spin-down age $\tau$; and that between the PWN size ($R$) and $\tau$.
While previous attempts of interpreting such correlations were assuming that they essentially reflect the evolution of a ``typical'' PWN, here I propose an explanation based on the assumption that the observed correlations originate as ``combined effects'' of PWNe evolving under a wide range of ambient conditions.
A rather unexpected consequence of this explanation is that in middle-age PWNe detected in the X-rays magnetic fields must be in general so weak that the evolution of the synchrotron X-ray emitting electrons is only weakly affected by synchrotron losses.
Finally, within the same framework, I will also present the observed statistical hardening of the PWN X-ray spectrum (Gotthelf 2003; Li, Lu \& Li 2008) with decreasing pulsar spin-down power $\Edot$ (corresponding to increasing $\tau$) as a possible effect of the superposition of two populations: young PWNe, with ages comparable with $\tauz$, and older ones, experiencing the reverberation phase.

\section{The observational evidence}

In this section I will briefly outline two important pieces of information, which came from statistical analyses of PWN samples.

\subsection{The ``evolution'' of the X-ray emission}

A clear correlation is present between $\LX$ and $\tau$ (or, alternatively, $\Edot$).
A very strong statistical correlation has been claimed by Mattana et al.~(2009, hereafter M+09).
Using a sample of 14 sources detected by H.E.S.S., these authors find an excellent correlation especially between $\LX$ and $\Edot$, namely:
\begin{equation}
\label{eq:MattanaCorrEdot}
\log_{10}\LX\!=\!(33.8\pm0.04)+(1.87\pm0.04)\log_{10}\Edot_{37}
\end{equation}
(where $\Edot_{37}$ indicates $\Edot$ expressed in units of $10^{37}\U{erg\,s^{-1}}$).
They also find a very good correlation with $\tau$, namely:
\begin{equation}
\label{eq:MattanaCorrtau}
\log_{10}\LX\!=\!(33.7\pm0.04)-(2.49\pm0.06)\log_{10}\tau_{4}
\end{equation}
(where $\tau_{4}$ is in units of $10^{4}\U{yr}$).
On the contrary, they show that the TeV PWN luminosity does not present any clear trend with either $\Edot$ or $\tau$.

Other authors also find similar correlations  between $\LX$ and $\tau$, or $\Edot$, even if not so strong.
For instance, Vink, Bamba \& Yamazaki~(2011), using a dataset of about 40 objects (namely the list given by Kargaltsev \& Pavlov 2008, with some minor changes), also identify a correlation, even if not so strong as in M+09.
Instead, they claim to have measured two separate slopes: respectively $1.91\pm0.33$ for younger objects (namely with $\tau\leq1.7\E{4}\U{yr}$) and $1.29\pm0.20$ for older ones (with $\tau>1.7\E{4}\U{yr}$): in a sense, the intermediate slope given by M+09 could be justified by the fact that they considered younger and older objects together.

\subsection{The X-ray evolution of PWNe in size}

Another interesting piece of information comes from the analysis of Bamba et al.~(2010, hereafter B+10).
These authors carried on, with the Suzaku X-ray telescope, a search for very low surface brightness emission around pulsars with ages up to $\sim10^5\U{yr}$.
The most unexpected result that they found can be summarized, with their own words, as follows: ``A systematic study of a sample of eight of these PWNe, together with Chandra data sets, has revealed that the nebulae keep expanding up to 100 kyr''.

Even though these measurements are affected by large uncertainties, B+10 notice an intrinsic difficulty to account theoretically for such large synchrotron X-ray sizes (up to $\sim20\U{pc}$), if the relativistic electrons are injected close to the pulsar position, namely at the wind termination shock.
The formula for the lifetime of electrons emitting synchrotron X-rays with energy $\epssyn$ is in fact:
\begin{equation}
\label{eq:synchtimescale}
\tsyn=2\parfrac{B}{10\U{\mu G}}^{-3/2}\parfrac{\epssyn}{1\U{keV}}^{-1/2}\U{kyr}.
\end{equation}
This means that the X-ray emitting region must be confined to the very inner regions of the PWN, unless the magnetic field is very low, in the $\mu G$ range.
B+10 then reach the following conclusions:
\begin{itemize}
\item[--]
$B$ must decrease in time down to very low values;
\item[--]
electron diffusion must be important, in order to justify a timescale for the escape smaller than the PWN age;
\item[--]
even under the above assumptions, a steadily increasing X-ray size with age, up to large extents for old PWNe, is anyway surprising. One would have expected instead that older PWNe have already been crushed by the supernova remnant (SNR) reverse shock, then decreasing their size.
\end{itemize}

\section{Essentials for modeling the PWN evolution}

This section is aimed at providing some basic mathematical tools to support the arguments that I will present in Sect.~4.

One-zone numerical models, in which homogeneous conditions are assumed within the PWN, are usually adopted to trace the evolution of a PWN+SNR system and of the PWN emission in the free-expansion phase and beyond (e.g.\ Gelfand, Slane \& Zhang~2009, hereafter GSZ09; Bucciantini, Arons \& Amato~2011).
From them one may infer that there are two epochs in the PWN evolution in which it may become particularly bright in various spectral ranges, including the X-rays.
Firstly, when the PWN is very young and, secondly, during the so-called ``reverberation phase'', when the PWN is hit by the SNR reverse shock.
During this phase, the PWN experiences a considerable decrease of its size: this not only implies an adiabatic energization of electrons, but also an increase of the nebular field, because of magnetic flux conservation.

Numerical models can follow the evolution with much higher accuracy than analytic models, but suffer of the intrinsic limitation that they can be computed only for a finite number of parameter choices; while, in order to extract correlations to be compared with the observed ones, one should in principle model the PWN+SNR evolution for all possible cases, weighting them with their respective probability of detection.
For this reason, in this section I will introduce an analytic model, or better some concepts and approximations that are behind analytic models, which can help us to get a better understanding of the role of the various parameters.
In the formulae I will also drop all numerical factors, which would be anyway inaccurate, and I will focus just on the dimensional dependencies.

The main parameters entering into a PWN+SNR model are:
the energy of the supernova ($\ESN$) and the mass of ejecta ($\Mej$);
the initial spin-down power ($\Edotz$) and the initial spin-down time ($\tauz$);
the conversion efficiencies with which electrons ($\etae$) and magnetic fields ($\etaB$) are injected into the PWN;
the ambient medium density ($\rhoISM$, in mass; or $\nISM$ in number). 

Let me first schematize the pulsar evolution in two phases: an earlier one (for $t<\tauz$), in which the pulsar power keeps constant (equal to $\Edotz$), while the PWN expands pushing against the supernova ejecta; and a later one (for $t>\tauz$), when the PWN freely expands in the cavity formed in the ejecta during the the former phase, while the magnetic field evolves conserving approximately its flux.
The former PWN phase can be described by the following dimensional equations:
\begin{eqnarray}
\label{eq:phase1eq1}
\etaP\Edotz t&\sim&\ESN\parfrac{\Mej\RPWN^2}{\ESN t^2}^{5/2};	\\
\label{eq:phase1eq2}
\etaB\Edotz t&\sim&\BPWN^2\RPWN^3\sim\Phi^2/\RPWN,\end{eqnarray}
where $\etaP$ is the injection efficiency in the components that then contribute to the pressure, namely the magnetic field and the particles with minor synchrotron energy losses ($\etaB<\etaP<\etaB+\etae$).
Eq.~\ref{eq:phase1eq1} tells us that a fraction of the total energy released by the pulsar goes into kinetic energy of the swept up ejecta (here a flat ejecta density profile is assumed); while Eq.~\ref{eq:phase1eq2}  that the total injected magnetic energy must be comparable with the present magnetic energy, and also defines the magnetic flux $\Phi$.
Note that, for indicating the PWN size, I use $\RPWN$ when I consider the actual evolution of an individual PWN, while simply $R$ for the values measured, which enter into the correlations.
The latter phase ($t>\tauz$) can be described instead by:
\begin{eqnarray}
\label{eq:phase2eq1}
\RPWN(t)&\sim&\Vz t\;			\nonumber\\
&&\!\!\!\!\!\!\!\!
\hbox{where: }
\Vz\sim\parfrac{\etaP^2\Edotz^2\tauz^2\ESN^3}{\Mej^5}^{1/10}\!\!\!\!;	\\
\label{eq:phase2eq2}
\BPWN(t)&\sim&\frac{\Phiz}{\RPWN^2}\;			\nonumber\\
&&\!\!\!\!\!\!\!\!
\hbox{where: }
\Phiz\sim\parfrac{\etaP^2\etaB^{10}\Edotz^{12}\tauz^{22}\ESN^3}{\Mej^5}^{1/20}\!\!\!\!.
\end{eqnarray}
The quantities $\Vz$ and $\Phiz$ are respectively the PWN expansion velocity and magnetic flux at $\tauz$ and beyond, and their formulae are derived from Eqs.~\ref{eq:phase1eq1} and \ref{eq:phase1eq2}, for $t=\tauz$.
The constancy of $\Vz$ and $\Phiz$ is only approximately true: for instance, the asymptotic constancy of the magnetic flux is valid only for braking indices $n>3$.
Since for simplicity I will use here $n=3$ (magnetic dipole-like braking), corresponding to the pulsar power evolution
\begin{equation}
\label{eq:Edotevol}
\Edot=\frac{\Edotz}{(1+t/\tauz)^2},
\end{equation}
in the case of linear expansion the magnetic flux will continue to increase logarithmically.
These formulae are intended to hold until the arrival of the SNR reverse shock, at the beginning of the Sedov phase, occurring nearly at
:
\begin{equation}
\label{eq:begSedovtime}
\tc\sim\parfrac{\Mej^5}{\rhoISM^2\ESN^3}^{1/6}
\end{equation}
(I assume $\tauz\ll\tc$).
The formula for $\tc$ has been obtained combining the requirements that the supernova energy has been converted into the motion of the ejecta ($\ESN\sim\Mej R^2/t^2$) and that the swept up mass of the ambient medium is comparable with the mass of the ejecta ($\Mej\sim\rhoISM R^3$).

The reverberation phase is not instantaneous, so I assumed it to take place from a time $\tcm$ to a time $\tcp$, both of the order of the characteristic time $\tc$.
The PWN size and magnetic field at the beginning of the reverberation phase are then:
\begin{eqnarray}
\label{eq:RPWNtcm}
\RPWN(\tcm)&\sim&\Vz\tc\qquad\qquad\quad\;\propto\rhoISM^{-1/3};	\\
\label{eq:BPWNtcm}
\BPWN(\tcm)&\sim&\Phiz/\RPWN(\tcm)^2\propto\rhoISM^{2/3}.
\end{eqnarray}
The PWN compression phase, during which the magnetic flux keeps constant, ends only when the pressure into the nebula will roughly balance that in the Sedov-like SNR.
This implies:
\begin{eqnarray}
\label{eq:BPWNtcp}
\BPWN(\tcp)&\sim&\parfrac{\etaB\rhoISM\ESN}{\etaP\Mej}^{1/2}\propto\rhoISM^{1/2}\\
\label{eq:RPWNtcp}
\RPWN(\tcp)&\sim&\parfrac{\Phiz}{\BPWN(\tcp)}^{1/2}\;\propto\rhoISM^{-1/4}.
\end{eqnarray}
It should be noticed that, under typical conditions, at least at the beginning of the reverberation phase the PWN field is so low that the energy evolution of the X-ray emitting electrons is still dominated by the adiabatic losses; and, during the PWN compression, the adiabatic term will then turn from losses to energy gains.
In the meanwhile, the PWN magnetic field will increase with compression (at constant magnetic flux), so that synchrotron losses will become more and more important.
If $\BPWN$ will become sufficiently strong, the evolution of the X-ray emitting electrons may turn from adiabatically to synchrotron dominated; otherwise, synchrotron losses can be neglected at all times, even at $\tcp$.

Let me put this in a more quantitative (thought still approximated) way.
I will start from the very basic equations for synchrotron emission:
\begin{eqnarray}
\label{eq:synch1}
-\frac{d\gm}{dt}&=&c_1B^2\gm^2,\;\,\hbox{where: }c_1=1.29\E{-9}\U{cgs};	\\
\label{eq:synch2}
\nusyn&=&c_2B\gm^2,\quad\hbox{where: }c_2=1.22\E{6}\U{cgs},
\end{eqnarray}
which give, respectively, the evolution of an electron Lorentz factor $\gm$ due to synchrotron losses and the characteristic frequency ($\nusyn$) at which this electron emits (under monochromatic approximation for the synchrotron emission).
The adiabatic and synchrotron time scales are then defined respectively as:
\begin{eqnarray}
\label{eq:tad}
\tad&=&\frac{\RPWN}{|\RPWNdot|};	\\
\label{eq:tsyn}
\tsyn&=&\frac{1}{c_1\BPWN^2\gm}.
\end{eqnarray}
We can estimate:
\begin{equation}
\label{eq:tadb}
\tad\sim\RPWN\,\sqrt{\frac{\Mej}{\ESN}},
\end{equation}
since the dimensional scaling for the reverse shock velocity is similar to that of the initial supernova velocity, namely $\sqrt{\ESN/\Mej}$.
Moreover, from Eq.~\ref{eq:tsyn} we get:
\begin{equation}
\label{eq:tsynb}
\tsyn\sim\frac{\RPWN^3}{c_1}\sqrt{\frac{c_2}{\nusyn\Phiz^3}}.
\end{equation}
Let me define $\teq$ as the time at which the adiabatic and synchrotron terms become comparable for a given $\nusyn$.
Taking $\tad=\tsyn$ leads to:
\begin{equation}
\label{eq:Reqsol}
\RPWN(\teq)\sim\parfrac{c_1^2\nusyn\Phiz^3\Mej}{c_2\ESN}^{1/4}.
\end{equation}
An important feature of this formula is that, differently from Eqs.~\ref{eq:RPWNtcm} and \ref{eq:RPWNtcp}, here the PWN size does not depend at all on $\rhoISM$.

Finally, the requirement that electrons emitting at $\nusyn$ do actually experience a transition from adiabatic to synchrotron-dominated evolution, namely that $\RPWN(\teq)>\RPWN(\tcp)$, leads to the condition:
\begin{equation}
\label{eq:teqcondition}
\frac{c_1^2\etaB\nusyn\rhoISM\Phiz}{c_2\etaP}>1
\end{equation}
The consequences of these results will be discussed in the next section.

\section{Correlations as combined statistical effects}

The usually proposed explanations for the X-ray and TeV behaviors can be synthesized by the following sentence in M+09: ``the X-ray emission traces the recent history of the nebula, whereas the $\gm$-ray emission traces a longer history, possibly up to the pulsar birth''.
The underlying hypothesis is that the observed correlations trace the actual time evolution for a ``typical'' PWN.
This is in fact a reason why B+10 found surprising that the observed evolution of the PWN size with $\tau$ does not show any feature reminiscent of the PWN contraction at the reverberation time. 

What I propose here is a different and I believe more natural kind of explanation, and I will use it to address the case of the X-ray emission (the $\gm$-ray emission will be treated elsewhere, and anyway it has already been discussed by some other authors; e.g.\ Tanaka \& Takahara 2013).
 The hypothesis at the basis of this explanation is that the observed correlations of both X-ray luminosity and size with age do not outline the evolution of a ``typical'' PWN; they originate instead as ``combined effects'' of PWNe evolving under different conditions.
The philosophy of this approach is similar to that we have already used to explain correlations in samples of radio SNRs (Bandiera \& Petruk 2010).

\subsection{A fictitious time dependence}

I will now combine some of the results presented in the previous section, with the aim of reproducing the time dependence of the PWN size, as observed by B+10.

Let me start with the following assumptions:
\begin{itemize}
\item[--]
The primary parametric dependence is that between $R$ and $\rhoISM$: this mostly because the ambient density is one of the parameters that vary most (over three orders of magnitude even without taking into account extreme conditions), while in many of the formulae given above the power-law dependences on $\rhoISM$ are not small when compared to those on other parameters.
\item[--]
On the other hand, the other two parameters which $\tc$ depends on, namely $\Mej$ and $\ESN$, are expected to change less (typically one order of magnitude or even less).
\item[--]
Aged PWNe are preferentially detected in the X-rays when, during the crushing phase, they reach their maximum X-ray brightness.
If the PWN size at the epoch of maximum brightness depends on $\rhoISM$, secondary correlations with $t$ and $\Edot$ will then follow from this primary correlation  with $\rhoISM$.
\end{itemize}
In fact, using Eqs.~\ref{eq:Edotevol} and \ref{eq:begSedovtime}, one may rewrite $\rhoISM$ as either a function of $t$ or of $\Edot$:
\begin{equation}
\label{eq:translaterhoISM}
\rhoISM\sim\parfrac{\Mej^{5/2}}{\ESN^{3/2}}t^{-3}\sim\parfrac{\Mej^{5/2}}{\ESN^{3/2}\Edotz^{3/2}\tauz^3}\Edot^{3/2}
\end{equation}
so that if $R\propto\rhoISM^{-\bt}$, the secondary relations $R\propto t^{3\bt}$ and $R\propto t^{-3\bt/2}$ will also follow.
These power-law relations will therefore be independent of the expansion evolution of PWNe before the reverberation phase.
In addition, no change of slope has to be expected in the above mentioned correlations, as due to the SNR reverse shock compression; indeed, the dynamics of the reverberation itself enters into generating the observed power-law correlations.

\begin{figure}
\includegraphics[width=80mm]{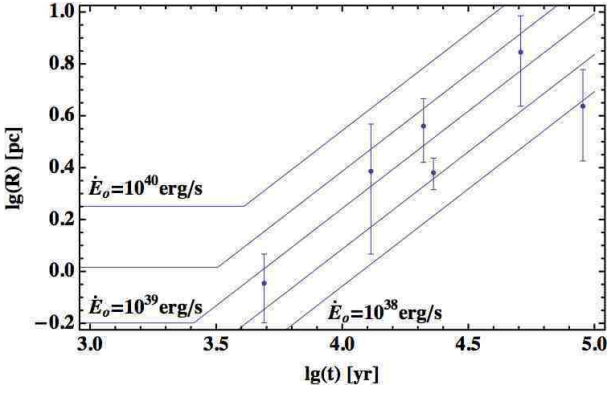}
\caption{\small Dependence of the PWN size on the pulsar spin-down age. Data from B+10 are compared with a set of analytical models, with $\Edotz$ made to vary from $10^{38}\U{erg\,s^{-1}}$ to $10^{40}\U{erg\,s^{-1}}$ , while the other parameters are the same as in GSZ09.}
\label{compareBambamoreEdots}
\end{figure}

The epoch at which a PWN is most likely detectable is, of course, that at which it reaches its maximum brightness.
If  $\teq<\tcp$, the PWN luminosity is keeping increasing until $\teq$ due to compression, while after then it rapidly decreases because the electrons with the required $\gm$ will quickly burn out: therefore in this case it would be most likely to detect the PWN close to $\teq$.
However, this scenario is not consistent with the observational evidence, since Eq.~\ref{eq:Reqsol} implies $\bt=0$, and therefore no correlation would be expected with either $t$ or $\Edot$. 
A different situation occurs if, instead, during the reverberation phase there is no adiabatic-to-synchrotron transition for the X-ray emitting electrons.
In this case the PWN will continue to brighten until $\tcp$, which then becomes the most likely epoch for detection.
Eq.~\ref{eq:RPWNtcp} would then lead to the following secondary correlation:
\begin{equation}
\label{eq:Rfictevol}
R\sim\parfrac{\etaP^2\Phiz^2\ESN}{\etaB^2\Mej^3}^{1/8}t^{3/4}.
\end{equation}
The results are shown in Fig.~\ref{compareBambamoreEdots}, in comparison with the data from B+10.
Note that I have deleted from the original data the two youngest objects, Kes~75 and MSH~15--52, because they are known not to have reached the reverberation phase yet.
Moreover, since B+10 have conventionally defined the PWN size as 3 times the value that they derived for $\sg$ from a gaussian fit to the observed profiles, I have corrected back for that factor before comparison.

For most physical parameters here I use the values as given in the numerical model presented by GSZ09.
Among them are: $\ESN=10^{51}\U{erg}$, $\Mej=8\,\Msun$, $\tauz=500\U{yr}$, $\etae=0.999$, $\etaB=0.001$.
In addition, I have assumed $\etaP=0.1$ (a parameter not needed in the numerical model, because self-computed).
The various lines in Fig.~\ref{compareBambamoreEdots} correspond to values of $\Edotz$ ranging from $10^{38}\U{erg\,s^{-1}}$ to $10^{40}\U{erg\,s^{-1}}$ ; while each line is obtained by varying the ambient density.
The functional trends should be rather reliable, becoming asymptotically flat (as described by Eq.~\ref{eq:Reqsol}) at higher ambient densities, while they are $\propto t^{3/4}$ (according to Eq.~\ref{eq:RPWNtcp}) at lower ambient densities.
Since in the present treatment I do not keep track of the numerical factors in the formulae, the positioning of a line for a given set of parameters is rather coarse.
And, anyway, analytic treatments would not allow to reach substantially higher accuracies, unless they are combined with numerical models (Bandiera, in preparation).

\begin{figure}
\includegraphics[width=80mm]{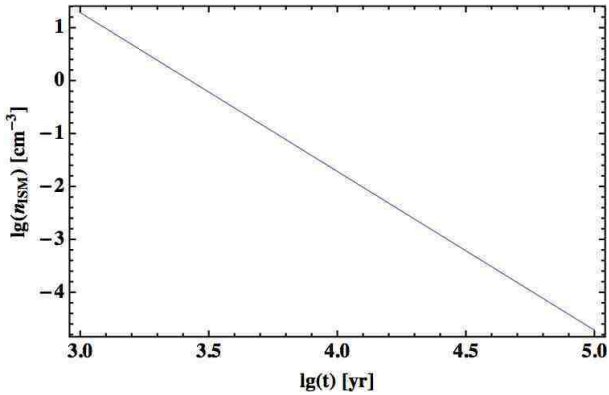}
\caption{\small Relation between the ambient medium density and the beginning of the Sedov phase (see Eq.~\ref{eq:begSedovtime}), for $\ESN=10^{51}\U{erg}$ and $\Mej=8\,\Msun$, plotted here as a fictitious function $\nISM(t)$. By using this relation, the primary dependence on $\nISM$ of $\RPWN$ at either $\teq$ or $\tcp$ can be turned into a ``fake'' dependence on $t$.}
\label{nISMoft}
\end{figure}

For the sake of illustration, let me draw here some comparisons between the results of the analytic model presented here and those of the GSZ09 model: my analytic estimates for $\tcp$, $\RPWN(\tcp)$ and $\BPWN(\tcp)$ are respectively a factor 0.29, 0.97, and 0.14 times the values found by their numerical model.
It is also worth plotting the ambient densities corresponding to the range of ages as in the B+10 sample.
The range of density values is centered to about $10^{-3}$--$1\U{cm^{-3}}$, a very natural one for the ISM medium (see Fig.~\ref{nISMoft}).

\subsection{The evolution of the X-ray luminosity}

Since in older PWNe the electrons now emitting X-rays did not suffer appreciable synchrotron losses for most of the time at earlier ages, predicting their synchrotron evolution is not difficult.
For a injection spectrum $\propto\gm^{-2}$, the energy distribution of the electrons at $t\simlt\tcm$ can be expressed as:
\begin{equation}
\label{eq:Nsolm}
N(\gm)\sim\frac{\etae\Edotz\tauz^2}{mc^2\Lmb t}\gm^{-2},
\end{equation}
where $\Lmb$ is a logarithmic factor, which evaluates $\sim20$.

Let me limit here the discussion to the case with slope $=2$ since it is easier, and usually consistent with the X-ray spectra.
A range of slopes at injection is however likely, and it will implies a broadening of the correlations.

The synchrotron luminosity of these electrons is:
\begin{equation}
\label{eq:Lsyn}
\Lsyn\sim\frac{c_1}{c_2}mc^2B\gm N(\gm),
\end{equation}
where $\gm$ is obtained from Eq.~\ref{eq:synch2}.
This can be rewritten as:
\begin{equation}
\label{eq:Lsynb}
\Lsyn\sim\frac{c_1\etae\Edotz\tauz^2\Phiz^{3/2}\Vz}{\Lmb\sqrt{c_2\nu}}\RPWN^{-4}.
\end{equation}
This equation holds not only before $\tcm$, but also within the reverberation phase, as long as the electrons considered evolve adiabatically.

Given Eq.~\ref{eq:RPWNtcm} it turns out that, for $\RPWN(\tcm)$, $\Lsyn\propto\rhoISM^{4/3}\propto\Edot^2$.
If $\teq<\tcp$, with the luminosity peak occurring at $\RPWN(\teq)$, $\LX$ would be independent of $\rhoISM$, and then statistically also of $\Edot$.
In the other case (namely that which already gives a radial evolution consistent with the data) the peak of luminosity would be near $t=\tcp$, so that (from Eq.~\ref{eq:RPWNtcp}) $\LX\propto\rhoISM\propto\Edot^{3/2}$.

The model prediction is in reasonably good agreement with the data: while M+09 obtain a significantly steeper trend (with slope $1.87\pm0.04$), for older PWNe Vink et al.~(2011) give in fact a shallower dependence (with slope $1.29\pm0.20$), just 1-$\sg$ away from 1.5, the value that my model predicts.

\begin{figure}
\includegraphics[width=80mm]{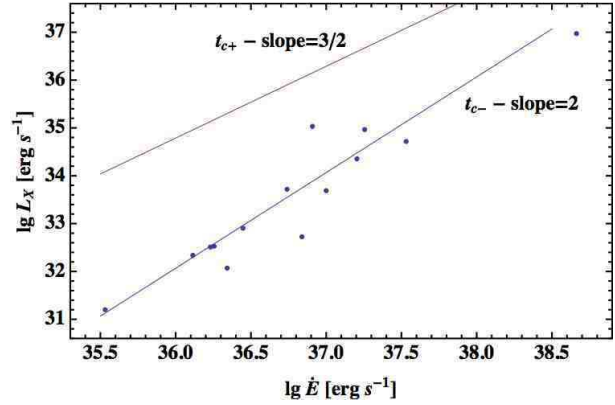}
\caption{\small Comparison of predictions from dimensional models with data from M+09. The line labelled as $\tcm$ (with slope 2) indicates the behavior of PWN X-ray luminosities when observed at the beginning of the reverberation phase, while the line labelled as $\tcp$ (with slope 3/2) gives their behavior at maximum compression. The  case $t=\tcp$, which I have shown to be in agreement with the observed $R(t)$ trend, seems reasonably consistent in slope also with the observed $\LX(\Edot)$. On the other hand, the case $t=\teq$ has again to be discarded, because it predicts a slope $=0$ (see text).}
\label{MattanaNoFitNumerical}
\end{figure}

\subsection{Consequences on the X-ray spectral index}

The framework that I have presented implicitly contains a prediction on the X-ray spectral index.
It is known that young X-ray PWNe have typically X-ray photon indices $\GmPWN\sim2$ (Asaoka \& Koyama 1990), corresponding to a spectral index $\sim1$.
In these PWNe the magnetic fields are rather strong, so that the evolution of X-ray emitting electrons is dominated by synchrotron losses: this spectral index then agrees with an injected electron distribution $\propto\gm^{-2}$.

On the other hand, during the evolution phases that I have considered above, the X-ray emitting electrons are either at the transition between an adiabatic to a synchrotron-dominated evolution (case $t=\teq$), or still evolving adiabatically (case $t=\tcp$).
In either case, the particle distribution function has been mostly built up while these electrons were evolving adiabatically, and therefore should have maintained the same spectral slope as at the injection.
This means that aged PWNe are most likely detected when they show a spectral index $\sim0.5$, i.e.\ $\GmPWN\sim1.5$.

In fact, Gotthelf (2003) investigated some statistical properties of the X-ray emission from 9 PWNe and associated pulsars showing, among others, evidence for an increase of $\GmPWN$ with $\Edot$ (which also means a decrease with the spin-down age).
A more recent work (Li et al.~2008), extending the analysis to 27 objects, confirmed the increase of $\GmPWN$ with $\Edot$ (see Fig.~6 in Li et al.~2008), namely that younger PWNe have typically harder X-ray spectra than the older ones.
Moreover, a plot of $\GmPWN$ against $\LX/\Edot$ (see their Fig.~7) shows a clear correlation between $\GmPWN$ and the PWN X-ray conversion efficiency.
That plot could be alternatively be interpreted as the combination of two populations of PWNe, one with $\GmPWN\sim2$ and larger $\LX/\Edot$, and the other with lower efficiencies and harder photon indices.
Also Kargaltsev \& Pavlov (2008) show a correlation between $\GmPWN$ and the PWN X-ray conversion efficiency (which in turn depends on the age).

All this seems at least qualitatively consistent with the idea that the PWN sample is composed of two populations: one of young PWNe, with typically $\GmPWN\sim2$, and the other of older ones, with typically $\GmPWN\sim1.5$.
This effect should be convolved with the actual distribution of injection spectrum slopes, but this is beyond the scope of the paper.
 
\section{Conclusions}

In this paper I have put forward a simple and I think reasonable working hypothesis, namely that observed statistical trends do not necessarily reflect details of the evolution of a ``typical'' PWN, but arise instead as a combined effect of PWNe evolving under different conditions.
Using analytic models and concepts, I have discussed some consequences of this hypothesis.
The resulting scenario is the following:
\begin{itemize}
\item[--]
Most of the middle-age PWNe are observed at or near the reverberation phase, when they are hit and squeezed by the SNR reverse shock.
\item[--]
The wide range of measured ages mostly depends on the fact that in different PWNe the reverberation phase occurs at different times. A primary parameter that rules that time is the ambient medium density (a quantity with order-of-magnitude variations from case to case).
\item[--]
At the beginning of the reverberation phase in all PWNe the magnetic field is so weak that even the X-ray emitting electrons evolve adiabatically. During the reverberation phase the PWN shrinks, the magnetic field increases, so that synchrotron losses may or may not become important for the evolution of the X-ray electrons.
 \item[--]
The statistical trends of the observed PWN sizes, as well as of their X-ray luminosities, can however be accounted for only if the X-ray electrons keep evolving adiabatically over the whole reverberation phase.
\item[--]
As a straightforward consequence, even if the energy distribution of the injected electrons does not change its slope with time, the X-ray spectral indices of older PWNe (namely of those near the reverberation phase) should be harder than those of young PWNe (namely of those with age not much larger than $\tauz$). This effect has been actually observed. 
\end{itemize}
The present work has however some main limitations.
First, not all possible physical cases have been taken into account: for instance, in the case that a strong turbulence is developed when the SNR reverse shock hits the PWN, one should reconsider the law according to which the PWN is compressed; and, also, an asymmetric compression of the PWN could be less effective.
Last but not least, the analytic (or, even worse, just dimensional) model presented here may be useful for outlining the slopes of functional dependences, but is very inaccurate in the determination of the absolute values of the various quantities.
In order to get more accurare estimates, it must be confronted with the results of a one-zone numerical model (Bandiera, in preparation).

\acknowledgements
I mostly thank Niccol\`o Bucciantini and Joseph Gelfand for interesting discussions and suggestions.
This work was partially funded through grant PRIN INAF 2010.


\end{document}